\documentclass[aps, twocolumn, superscriptaddress,longbibliography, amsfonts,amssymb,amsmath,a4paper,floats,floatfix]{revtex4-2}

\usepackage[utf8]{inputenc}
\usepackage[english]{babel}
\usepackage{graphicx}
\usepackage{dcolumn}
\usepackage{bm}
\usepackage{siunitx}
\usepackage{xspace}
\usepackage{xcolor, color, colortbl}
\usepackage{braket}
\usepackage{booktabs}
\usepackage{notes2bib}
\definecolor{LightGray}{gray}{0.9}	
\usepackage{lineno}

\newcommand{\vtg}{\ensuremath V_{\mathrm{TG}}}
\newcommand{\vbg}{\ensuremath V_{\mathrm{BG}}}   
 
\newcommand{\mb}[1]{\mathbf{#1}}

\DeclareSIUnit\angstrom{\text {\AA}}

\begin{document}

\title{Spin-orbit proximity in MoS$_2$/bilayer graphene heterostructures}

\author{Michele Masseroni}%
     \email{masmiche@phys.ethz.ch}
    \affiliation{Solid State Physics Laboratory, ETH Z\"urich, 8093 Z\"urich, Switzerland}
\author{Mario Gull}%
    \affiliation{Solid State Physics Laboratory, ETH Z\"urich, 8093 Z\"urich, Switzerland}
\author{Archisman Panigrahi}
    \affiliation{Department of Physics, Massachusetts Institute of Technology, Cambridge, MA 02139, USA}
\author{Nils Jacobsen}
    \affiliation{1st Physical Institute, Faculty of Physics, University of Göttingen, 37077 Göttingen, Germany } 
\author{Felix Fischer}
    \affiliation{Solid State Physics Laboratory, ETH Z\"urich, 8093 Z\"urich, Switzerland}
\author{Chuyao Tong}
    \affiliation{Solid State Physics Laboratory, ETH Z\"urich, 8093 Z\"urich, Switzerland}
\author{Jonas D. Gerber}
    \affiliation{Solid State Physics Laboratory, ETH Z\"urich, 8093 Z\"urich, Switzerland}
\author{Markus Niese}
    \affiliation{Solid State Physics Laboratory, ETH Z\"urich, 8093 Z\"urich, Switzerland}
\author{Takashi Taniguchi}
	\affiliation{Research Center for Materials Nanoarchitectonics, National Institute for Materials Science,  1-1 Namiki, Tsukuba 305-0044, Japan}
\author{Kenji Watanabe}
	\affiliation{Research Center for Electronic and Optical Materials, National Institute for Materials Science, 1-1 Namiki, Tsukuba 305-0044, Japan}
\author{Leonid Levitov}
    \affiliation{Department of Physics, Massachusetts Institute of Technology, Cambridge, MA 02139, USA} 
\author{Thomas Ihn}
    \affiliation{Solid State Physics Laboratory, ETH Z\"urich, 8093 Z\"urich, Switzerland}
\author{Klaus Ensslin}
    \email{ensslin@phys.ethz.ch}
    \affiliation{Solid State Physics Laboratory, ETH Z\"urich, 8093 Z\"urich, Switzerland}
\author{Hadrien Duprez}
    \affiliation{Solid State Physics Laboratory, ETH Z\"urich, 8093 Z\"urich, Switzerland}

\date{\today}

\begin{abstract}
Van der Waals heterostructures provide a versatile platform for tailoring electronic properties through the integration of two-dimensional materials. 
Among these combinations, the interaction between bilayer graphene and transition metal dichalcogenides (TMDs) stands out due to its potential for inducing spin-orbit coupling (SOC) in graphene. 
Future devices concepts require the understanding the precise nature of SOC in TMD/bilayer graphene heterostructures and its influence on electronic transport phenomena. 
Here, we experimentally confirm the presence of two distinct types of spin-orbit coupling (SOC), Ising ($\Delta_\mathrm{I}= \SI{1.55}{meV}$) and Rashba ($\Delta_\mathrm{R}= \SI{2.5}{meV}$), in bilayer graphene when interfaced with molybdenum disulphide, recognized as one of the most stable TMDs.
Furthermore, we reveal a non-monotonic trend in conductivity with respect to the electric displacement field at charge neutrality. This phenomenon is ascribed to the existence of single-particle gaps induced by the Ising SOC, which can be closed by a critical displacement field.
Remarkably, our findings also unveil sharp peaks in the magnetoconductivity around the critical displacement field, challenging existing theoretical models.
\end{abstract}

\maketitle


\section{Introduction}

Spin is emerging as a promising alternative or complement to charge for information storage and processing \cite{burkard_semiconductor_2023}. 
Spin-orbit coupling (SOC) is crucial in spin-based devices, enabling manipulation of spin states through time-dependent electric fields \cite{Rashba_orbital_2003, khoo_on-demand_2017}.
Bernal bilayer graphene (BLG) holds potential for spintronics \cite{avsar_colloquium_2020} and quantum computing \cite{trauzettel_spin_2007}, with recent studies indicating long spin relaxation times in BLG quantum dots \cite{gachtergarreis_single_2022, garreistong_long_2024, denisov_ultra-long_2024}. 
However, intrinsic Kane-Mele (KM) SOC \cite{kane_quantum_2005} in graphene is weak ($\SIrange{40}{80}{\micro eV}$) \cite{sichau_resonance_2019, duprez_spectroscopy_2023}. 
Various methods have been explored to enhance SOC in BLG, including interfacing with high-SOC substrates. 
Transition metal dichalcogenides (TMDs) have shown promise in this regard, offering significant SOC enhancements (from $\SIrange{1}{10}{\meV}$) without compromising graphene's electronic quality \cite{avsar_spinorbit_2014, wang_origin_2016, island_spinorbit_2019}.
Additionally, the combination of BLG on WSe$_2$ has recently shown to host an unexpected superconducting phase, where the SOC seems to play a major role \cite{zhang_enhanced_2023,holleis_ising_2023}, prompting further study of such heterostructures.

The extrinsic SOC induced in BLG by the TMDs is described by the Hamiltonian \cite{gmitra_proximity_2017}
\begin{equation}\label{eq:SO_hamiltonian}
    H_\mathrm{SO} = \frac{\Delta_\mathrm{I}}{2}  \xi s_z \mathbb{I}_\sigma + \frac{\Delta_\mathrm{R}}{2} (\xi\sigma_x s_y - \sigma_y s_x),
\end{equation}
where $\xi=\pm1$ represents the valley index, $s_{x,y,z}$ denote spin Pauli matrices, $\sigma_{x,y}$ and $\mathbb{I}_\sigma$ are Pauli and unit matrices operating on the sublattice degree of freedom $(A_1, B_1)$ within the layer in contact with the TMD [see schematic in Fig.~\ref{fig:figure3}(e)]. 
The first term, known as the Ising SOC, acts similar to an effective out-of-plane magnetic field with a valley-dependent sign.
It lifts the four-fold spin and valley degeneracy at the $K\pm$ points, forming spin-valley locked Kramers doublets.
The second term is a Rashba type of SOC \cite{rashba_graphene_2009}, favoring an in-plane spin polarization perpendicular to the sublattice isospin vector. 

Intensive theoretical \cite{gmitra_proximity_2017, khoo_tunable_2018, li_twist_2019, david_induced_2019, naimer_twist_2021, chou_enhanced_2022} and experimental efforts \cite{avsar_spinorbit_2014, wang_origin_2016, yang_strong_2017, zihlmann_large_2018, wang_quantum_2019, island_spinorbit_2019, sun_determining_2023} in understanding and quantifying SOC proximity effects, have led to a range of values for the relative strength of the two SOC terms depending on analysis method. 
This is because the strength of SOC is often inferred indirectly, for example, through the extraction of relaxation times obtained from quantum interference effects such as weak antilocalization (WAL) \cite{wang_strong_2015, wakamura_spin-orbit_2019, fulop_boosting_2021}.
In contrast, the fundamental frequency $f=\Delta(B^{-1})$ of the Shubnikov-de Haas oscillations (SdHOs) offers a direct measurement of the Fermi surface and is suitable for determining the band splitting induced by SOC \cite{bergemann_fermi_2005}.
However, the energy resolution of this technique is limited by the broadening of the Landau levels, necessitating high electron mobilities and low disorder potentials.

\begin{figure*}[tb]
    \centering
    \includegraphics{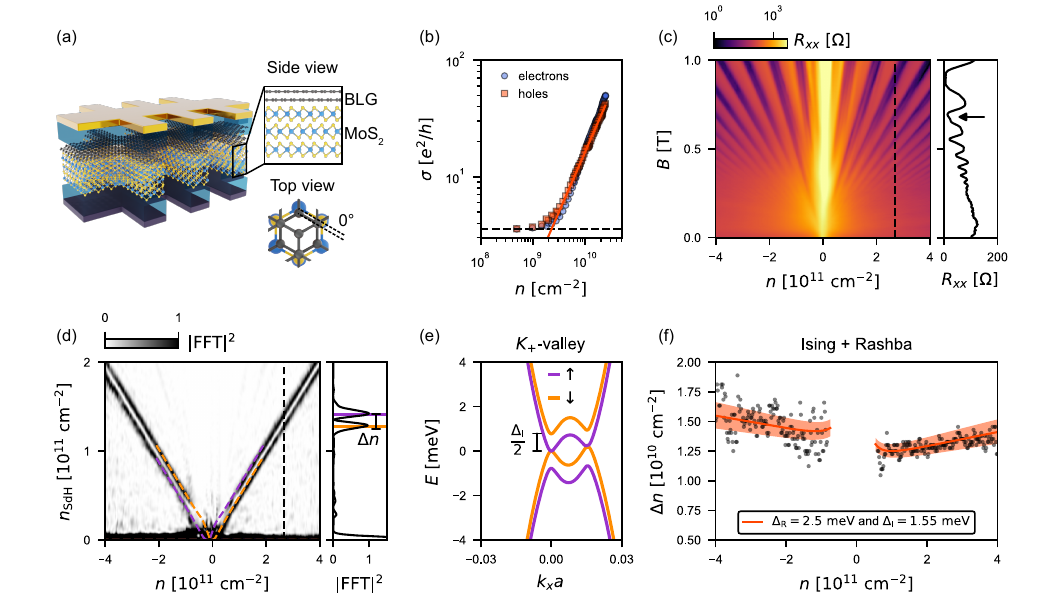}
    \caption{\textbf{Magnetotransport data at zero displacement field}. (a) Schematic representation of the BLG/MoS$_2$ heterostructure, illustrating a cross-section of the layers (top right panel) and highlighting the alignment of the BLG and MoS$_2$ layers (bottom right panel).
    (b) The conductivity is plotted as a function of carrier density on a logarithmic scale. The measurement was conducted at a temperature of approximately $\SI{30}{mK}$. Dotted (squared) markers represent data for electron (hole) doping. The solid red line depicts a linear fit, while the black dashed line indicates saturation of the conductivity.
    (c) Landau fan at zero displacement field (left panel) measured at a temperature $T=\SI{1.3}{K}$. The right panel displays a vertical linecut at $n=\SI{2.7E11}{cm^{-2}}$ (dashed line in the left panel). 
    (d) Fast Fourier transform (FFT) of the Landau fan shown in (c). The FFT of $R_{xx}(B^{-1})$ is calculated line-by-line for each density. The vertical axis has been rescaled according to $n_\mathrm{SdH}=2ef/h$, where $f$ is the frequency axis in Tesla, accounting for the valley degeneracy. Dashed lines represent densities obtained from the band structure in (e). 
    The right panel shows a vertical linecut at $n=\SI{2.7E11}{cm^{-2}}$ (dashed line in the left panel).
    (e) Band structure of bilayer graphene with SOC ($\Delta_\mathrm{I}=\SI{1.55}{meV}$ and $\Delta_\mathrm{R}=\SI{2.5}{meV}$) close to the $K_{+}$ point at zero displacement field. The bands are plotted along the relative momentum $k_x$ (or equivalently, along the line $\Gamma-K-M$ of the first Brillouin zone). The horizontal axis is scaled by the lattice constant $a=\SI{2.46}{\angstrom}$. The color of the bands encode the spin texture (violet for spin up and orange for spin down). 
    (f) Density difference $\Delta n$ obtained from the distance between the two peaks in the Fourier spectrum shown in (d). The red solid line represents a fit to the data, obtained from the band structure in (e) by determining the density of states and then the carrier densities $n_\mathrm{\downarrow, \uparrow}$ followed by calculating the difference $|n_\mathrm{\downarrow}-n_{\uparrow}|$.
    The fit yields the SOC parameters: $\Delta_{I}=\SI{1.55 \pm 0.1}{meV}$ and $\Delta_\mathrm{R}=\SI{2.5\pm 0.5}{meV}$. The shaded area indicates the uncertainty in the fitting parameters, reflected in the uncertainty in $\Delta n$.}
    \label{fig:figure1}
\end{figure*}

Here, we conduct magnetotransport experiments on a dual-gated MoS$_2$/BLG heterostructure.
First, we analyze SdHOs to quantify proximity-induced SOC.
Our results confirm the presence of both Ising ($\Delta_\mathrm{I}=\SI{1.55}{meV}$) and Rashba ($\Delta_\mathrm{R}=\SI{2.5}{meV}$) SOC.
Despite their comparable strength, we show that the splitting of the low energy bands mainly arises from the Ising SOC.
Additionally, we observe a non-monotonic conductivity response to an applied displacement field when BLG is charge neutral.
Our tight-binding calculations show how the displacement field $D$ opposes the Ising SOC, closing single-particle gaps in the spin-polarized bands at a critical value $D_\mathrm{c}$ and causing local maxima in the conductivity.
At this critical field, the application of an external magnetic field rapidly suppresses the conductivity, challenging existing theoretical models and suggesting the involvement of many-body interactions.

\section{Proximity induced spin-orbit coupling}

Determining the SOC gap in BLG via magnetotransport experiments is challenging due to disorder-induced density fluctuations $\delta n$. Shown in Figure~\ref{fig:figure1}(a) is the schematic of our sample, comprising BLG atop three layers of MoS$_2$, encapsulated within hexagonal boron nitride (hBN), and placed on a graphite bottom gate. 
The use of hBN dielectrics and a graphite layer minimizes density fluctuations \cite{rhodes_disorder_2019}, evident from the low density $\delta n\sim \SI{2E9}{cm^{-2}}$ at which conductivity saturates in our sample [Figure~\ref{fig:figure1}(b)]. High charge carrier mobilities [$\sim\SI{5E5}{cm^2(V~s)^{-1}}$ at $n=\SI{5E10}{cm^{-2}}$, see Supplementary Information] indicate minimal impact of the MoS$_2$ layer on BLG's electronic properties compared to hBN-encapsulated Bernal BLG devices \cite{yankowitz_van_2019}.

We analyze SdHOs at zero displacement field ($D$) and low magnetic fields ($B$) to determine the band splitting induced by the SOC. 
Figure~\ref{fig:figure1}(c) shows the longitudinal resistance $R_{xx}$ as a function of $B$ and electron density $n$ at $T=\SI{1.3}{K}$. 
Pronounced minima in resistance $R_{xx}$ occur at filling factors $\nu=\pm 4 N$ ($N$ an integer), characteristic of pristine BLG. 
In addition, small oscillation maxima appear in the SdH minima (highlighted by the arrow in the right panel), suggesting the presence of a broken symmetry.

To determine the oscillation frequency of the SdHOs, we employ a numerical fast Fourier transform (FFT) of $R_{xx}(1/B)$ calculated line-by-line for each density, as shown in Fig.~\ref{fig:figure1}(d). 
The FFT reveals two clear frequencies $f_{1}$ and $f_{2}$, resulting from the splitting of the Fermi surface, which is attributed to the influence of the MoS$_2$ substrate through the spin-orbit proximity effect. 
The sum of the electron densities $n_i=2ef_i/h$ ($i=1,2$) obtained from the SdHO matches the Hall density by accounting for the twofold valley degeneracy, as expected. 

The two SOC terms in equation~\eqref{eq:SO_hamiltonian} yield distinct density dependencies for the spin-orbit splitting. 
The Ising SOC induces a constant splitting as a function of the Fermi energy (and hence density), while the Rashba term leads to a splitting that increases with the Fermi energy. 
Although the splitting in Fig.\ref{fig:figure1}(d) initially appears constant with carrier density, a closer examination of $\Delta n$ in Fig.\ref{fig:figure1}(f) reveals a small but detectable slope. 
By aligning the density difference $\Delta n$ obtained from the tight-binding model (see Methods \ref{met:Tight_binding} and \ref{met:Fitting}) with the data (illustrated by the red solid line), we find $\Delta_\mathrm{I}=\SI{1.55\pm 0.1}{meV}$ and $\Delta_\mathrm{R}=\SI{2.5\pm 0.5}{meV}$ 
\bibnote{We acknowledge that the numerical outcome of the fit can be subtly influenced by the choice of the tight-binding intralayer and interlayer coupling parameters of BLG, referred to as the Slonczewski--Weiss--McClure parameters ($\gamma_0$ and $\gamma_1$). These parameters dictate the curvature of the bands, thereby affecting the conversion between energy and density, as elaborated in detail in Methods~\ref{met:Fitting}.
In addition, we chose to neglect the parameter $\gamma_4$, which governs the particle-hole asymmetry. This decision was made because including $\gamma_4$ would introduce an offset between the electron and hole doping that is not observed in our experiments.}.
Despite the similar magnitudes of the two SOC terms, the weak density dependence exhibited by $\Delta n$ suggests that the primary contribution to the splitting of low-energy states stems from the Ising SOC (see also discussion in Supplementary Information).
The theoretically predicted densities with these parameters are overlaid against the total density in Fig.~\ref{fig:figure1}(d) as orange and violet dashed lines, demonstrating good agreement with the experimental data.

We validate our findings at finite displacement fields, leveraging the layer-dependent SOC induced by the asymmetric structure of our sample \cite{khoo_on-demand_2017}. This layer selectivity is demonstrated in Extended Data \ref{fig:Layer_polarization}, where the electron wave function is polarized via the applied displacement field in one layer or the other, depending on its sign.
A detailed discussion of these data can be found in the Supplementary Information.

Next we continue the discussion by investigating the impact of SOC on the electrical conductivity ($\sigma$) of BLG at charge neutrality (CN).

\section{Conductivity at charge neutrality}\label{sec:CN_data}

\begin{figure*}[tb]
    \centering
    \includegraphics{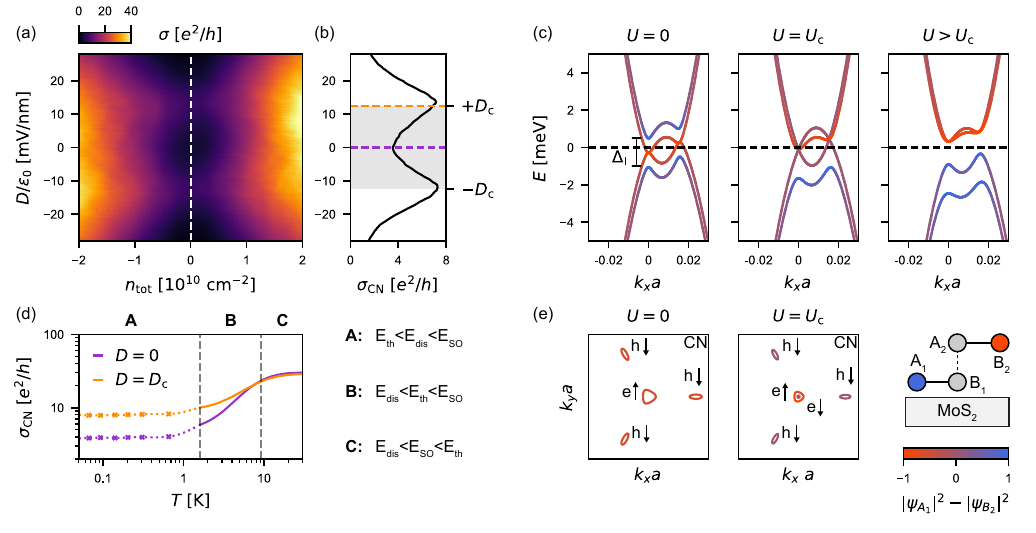}
    \caption{\textbf{Displacement field dependence of the conductivity at CN.} (a) Conductivity $\sigma$ as a function of density and displacement field measured at $T\approx\SI{30}{mK}$. 
    (b) Verical line cut of $\sigma$ in (a) at CN.  
    (c) Band structure of proximitized BLG at the $K$ point. The calculation include an Ising SOC term with $\Delta_\mathrm{I}=\SI{1.55}{meV}$. The band structure is shown for three characteristic interlayer potential energies: $U=0$ (left panel), $U=\SI{0.775}{meV}$ (central panel) and $U=\SI{2}{meV}$ (right panel). 
    The energy axis is adjusted such that $E=0$ corresponds to charge neutality, which is marked by the horizontal dashed lines. The color code represents the layer polarization: blue indicates polarization on layer 1, while red on layer 2, shown in the schematics in the bottom right panel.
    The band structure shown in the left panel is the same as Fig.~\ref{fig:figure1}(c), where we color coded the bands according to the spin texture.
    (d) Conductivity at CN for $D=0$ (violet) and $D=D_\mathrm{c}$ (orange) as a function of temperature in logarithmic scale. The crosses were measured in a dilution refrigerator, while the solid line was measured in a pumped He4 cryostat.
    (e) Constant energy contours of the Fermi energy at charge neutrality for $U=0$ (left panel) and $U=U_\mathrm{c}$ (right panel). The Fermi pockets are labelled according to their doping, electron $e$ and holes $h$, and their spin ($\uparrow, \downarrow$).
    }
    \label{fig:figure3}
\end{figure*}

Measurements of $\sigma$ reveals a non-monotonic dependence on the displacement field [Fig.~\ref{fig:figure3}(a)], which appears in a narrow density range ($n\sim\SI{1E10}{cm^{-2}}$) around CN.
A local minimum at $D=0$ is surrounded by conductivity maxima at $D=\pm D_\mathrm{c}\approx\SI{12.5}{mV/nm}$, as highlighted in the line cut at $n=0$ presented in Fig.~\ref{fig:figure3}(b).

This dependence can be understood by taking into account the influence of SOC on the BLG band structure.
From tight-binding calculations we find that the Rashba SOC has little effect on the low energy bands
(see Supplementary Information for more details).
For this reason, we consider only the Ising SOC in the following discussion.
The outcome of the band structure calculations is presented in Fig.~\ref{fig:figure3}(c), shown for the $K_{+}$-valley and three characteristic interlayer potential energies ($U$).
First, we consider the case $U=0$ in the left panel.
We observe that the band structure comprises two pairs of bands, one split by the energy $\Delta_\mathrm{I}$ and partially layer-polarized on the bottom layer (blue), while the other pair is degenerate at two points along $k_x$ and is partially polarized on the top layer (red).
Due to their partial layer polarization, the application of a displacement field shifts the two pairs of bands relative to each other based on their layer polarization.
Notably, the calculations show that a band gap emerges only once the interlayer potential energy exceeds the critical value $U_\mathrm{c}=\Delta_\mathrm{I}/2\approx\SI{0.8}{meV}$ (right panel), i.e. once $U$ counteracts the SOC, resulting in the closure of the gap between bands with the same spin [see Fig.\ref{fig:figure3}(e) for the spin texture]
\bibnote{
A recent experiment by Seiler et al. \cite{seiler_probing_2023} in high-quality BLG demonstrated that trigonal warping causes the overlap (approximately $\sim\SI{1}{meV}$) of conduction and valence bands, resulting in a semi-metallic phase. Due to the band overlap, a finite displacement field is necessary to suppress the density of states at charge neutrality (CN) and open the band gap. However, this effect does not lead to a non-monotonic dependence of the conductivity, as observed in our experiment.}.
This elucidates why in the experiment, the conductivity starts decreasing with increasing displacement fields only when $D>D_\mathrm{c}$, and associate the maxima in the conductivity with the closure of the SOC gaps.
We verify this interpretation by converting $U_\mathrm{c}$ into a displacement field, taking into account interlayer screening (see Methods~\ref{met:Screening} for details).
The conversion yields a displacement field of $\SI{11.4}{mV/nm}$, in good agreement with the experimental value of $D_\mathrm{c}\approx\SI{12.5}{mV/nm}$.
Furthermore, the local conductivity minimum at $D=0$ is observed only at low density and vanishes around $n\sim\SI{1E10}{cm^{-2}}$, which corresponds to the density required to fill the spin-orbit splitting of the bands, as demonstrated in Fig.~\ref{fig:figure1}(f).

The local minimum in conductivity at $D=0$ prompts consideration of a potentially insulating phase arising from the presence of a gap, as reported for BLG fully encapsulated in TMDs \cite{island_spinorbit_2019}.
To verify this, we examine the temperature dependence of the conductivity in Fig.\ref{fig:figure3}(d). 
Over the temperature range of $\SIrange{1}{10}{K}$, the conductivity increases by one order of magnitude, indicative of insulating behavior. 
However, the data only conforms to the Arrhenius law within a very limited temperature range (shown in Extended Data \ref{fig:Arrhenius}) and saturates to rather large conductivity values at low temperatures.  
Similarly, the conductivity at the critical field $D_\mathrm{c}$ also increases with temperature. 
Thus, although the insulating behavior is affected by the applied displacement field, it is consistently observed across all displacement fields at CN.

To further understand the temperature dependence, we compare the thermal energy $E_\mathrm{th}=k_\mathrm{B} T$ with the other characteristic energy scales determined by disorder ($E_\mathrm{dis}$) and SOC ($E_\mathrm{SO}$). 
First, we take into account the disorder potential, which induces density fluctuations of the order $\delta n\approx\SI{2E9}{cm^{-2}}$. 
These fluctuations are converted into an energy scale $E_\mathrm{dis}\approx\SI{0.14}{meV}$ using an effective mass approximation ($m^*\approx 0.035~m_0$, where $m_0$ is the bare electron mass \cite{mccann_electronic_2013}) and taking into account the twofold valley degeneracy.
At low temperatures ($E_\mathrm{th}<E_\mathrm{dis}$), the conductivity is governed by disorder-induced electron-hole puddles, causing the saturation of the conductivity in the temperature range labeled A in Fig.~\ref{fig:figure3}(d).
Second, the SOC introduces gaps $E_\mathrm{SO}=\Delta_\mathrm{I}/2\approx\SI{0.8}{meV}$ between bands of the same spin at $D=0$, as illustrated in Fig.\ref{fig:figure1}(c). 
The presence of these ``spin-resolved'' gaps, even without a real band gap, could explain the insulating behavior of the conductivity.
Effectively, if spin is conserved in thermal activation processes, carriers cannot be thermally excited from the highest occupied valence band into the lowest unoccupied conduction band, because these bands have opposite spin.
Therefore, carriers thermally excited above the spin gap $E_\mathrm{SO}$ should result in an increase of conductivity with increasing temperature, which is precisely happening in the temperature range labeled B in Fig.~\ref{fig:figure3}(d).
In regime C ($E_\mathrm{dis}<E_\mathrm{SO}<E_\mathrm{th}$), the thermal energy surpasses the SOC gap, causing the conductivity to saturate again.

Based on the results presented above, we attribute the dependence of conductivity on displacement field, density and temperature to the presence of spin-orbit-induced gaps in the spin-polarized bands in the absence of a global band gap.

\begin{figure}[tb]
    \centering
    \includegraphics[width=\columnwidth]{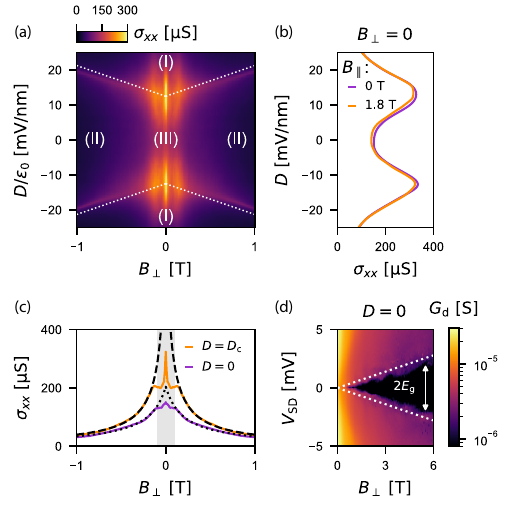}
    \caption{\textbf{Magnetotransport data at CN}. (a) Conductivity as a function of $B_\perp$ and $D$ at $T=\SI{30}{mK}$. This plot represents the $(D,B)$ phase diagram of SOC proximitized BLG. Phase (I) is a gapped phase with layer polarized wave function. Phase (II) is another gapped phase that has been attributed to the canted antiferromagentic phase in pristine BLG \cite{kharitonov_canted_2012, maher_evidence_2013}. The phase boundary between phase (I) and (II) (straight dotted lines) are described in Supplementary Information. 
    (b) $\sigma_{xx}$ plotted against $D$ for $B_\perp=0$ and two values of in-plane magnetic field: $B_\parallel=0$ (violet) and $B_\parallel=\SI{1.8}{T}$ (orange).
    (c) Linecuts of panel (a) at $D=0$ (violet) and $D=D_\mathrm{c}$ (orange). The black dashed and dotted lines highlight the $1/B$ dependence of the conductivity. 
    (d) Differential conductance $G_\mathrm{d}=\mathrm{d} I/\mathrm{d}V_\mathrm{SD}$ as a function of the voltage bias $V_\mathrm{SD}$ and magnetic field $B_\perp$ at $D=0$. The dotted line highlights $E_\mathrm{g}\propto B_\perp$.}
    \label{fig:figure4}
\end{figure}

\section{$(B, D)$ phase diagram}

In the final section of this work, we describe magnetotransport measurements at CN.
Figure~\ref{fig:figure4}(a) illustrates the longitudinal conductivity $\sigma_{xx}$ as a function of out-of-plane magnetic field ($B_\perp$) and displacement field at a temperature of $T\approx\SI{30}{mK}$. 
With the exception of the low magnetic field peaks at $D_\mathrm{c}$, which we discuss below, the phase diagram depicted in Fig.~\ref{fig:figure4}(a) bears resemblance to that of pristine BLG \cite{nandkishore_flavor_2010, knothe_phase_2016}. 
Drawing on previous studies \cite{weitz_broken-symmetry_2010, nandkishore_flavor_2010}, we partition the parameter space into three distinct regions. 

Phases (I) and (II), occurring at large displacement and magnetic fields respectively, are anticipated to mirror the behavior of the BLG system in the absence of SOC. This is attributed to the dominance of energy scales dictated by the externally applied parameters ($B_\perp$ and $D$) over the SOC gap $\Delta_\mathrm{I}/2$.
Hence, we attribute phase (I) to the layer-polarized insulating state arising from the band gap induced by the displacement field, as illustrated in Fig.~\ref{fig:figure3}(c).
Phase (II) represents the insulating state of the quantum Hall $\nu=0$ state.
In this phase, our bias spectroscopy measurements uncover the presence a gap $E_\mathrm{g}\propto B_\perp$ [Fig.\ref{fig:figure4}(d)], which qualitatively explains the observed $B_\perp^{-1}$ suppression of the conductivity [see dotted and dashed lines in Fig.\ref{fig:figure4}(c)]. 
This behavior aligns with the canted antiferromagnetic phase observed in pristine BLG \cite{kharitonov_antiferromagnetic_2012}.
Moreover, the boundaries between Phase (I) and (II) (indicated by white dotted lines and elaborated in detail in Supplementary Information) exhibit common characteristics with those observed in pristine BLG: the insulator-insulator transition features enhanced conductance \cite{weitz_broken-symmetry_2010, maher_evidence_2013}, and the displacement field required to induce the transition is $D^*(B)\propto ge^2B/2h$ \cite{nandkishore_flavor_2010, gorbar_energy_2010, toke_intra_2011, kharitonov_antiferromagnetic_2012}.
On the other hand, Phase (III), emerging at $B=0$ and $D=0$, is expected to differ from BLG samples not in proximity with a TMD layer.
Single-particle band structure calculations reveal that this phase is characterized by spin-polarized bands [Fig.~\ref{fig:figure1}(e)] with partially layer-polarized wave functions [left panel in Fig.~\ref{fig:figure3}(c)].
As we have discussed in Sec.\ref{sec:CN_data}, this phase shows thermal activation due to the presence of spin-orbit gaps in the spin-polarized bands without the presence of a global band gap, as illustrated by Fig.~\ref{fig:figure4}(d).
While the magnetic field induces a gap in the spectrum, no distinct phase transition is observed between Phase (III) and Phase (II). 
This is likely due to both phases being overall layer unpolarized when considering the occupied valence bands, as demonstrated by the phase transition induced by the displacement field.

Now, we examine the sharp magnetoconductivity peaks at $D_\mathrm{c}$ [see orange curve in Fig.~\ref{fig:figure4}(c)], a novel feature of spin-orbit proximitized BLG not previously reported. With current theoretical models unable to fully explain these peaks, we explore various possibilities.

At first glance, the sharp peak in the orange curve in Fig.\ref{fig:figure4}(c) resembles the signature of WAL, expected in materials with strong SOC. 
This effect has been observed in numerous transport experiments in SOC-proximitized graphene \cite{wang_strong_2015, wakamura_strong_2018, zihlmann_large_2018, wakamura_spin-orbit_2019, wang_origin_2016, tiwari_electric_2021, amann_counterintuitive_2022}. 
However, with a mean-free-path exceeding $\SI{1}{\micro m}$ at finite density, the condition $\ell_\phi > \ell_e$ (where $\ell_\phi$ and $\ell_e$ represent the phase-coherence length and mean-free-path, respectively) required to observe this effect would never be fulfilled ($\ell_\phi \leq \SI{360}{nm}$ if fitting the peak with a WAL model, as detailed in the Supplementary Information).
Furthermore, the magnitude of the peak ($\sim 3-4 e^2/h$) exceeds what would be expected for WAL, which typically reaches up to $0.5 e^2/h$ per conducting channel. 
Additionally, quantum interference effects are typically suppressed with increasing temperature due to the decrease in $\ell_\phi$. 
In contrast, the magnitude of the peak in $\sigma_{xx}$ remains robust against temperature changes [see \ref{fig:Tdependence_Peak} in Extended Data].
For these reasons, we conclude that the peaks cannot arise from WAL.

Typically, distinguishing between how a magnetic field affects orbital or spin degrees of freedom involves tilting the field with respect to the plane. Orbital effects couple exclusively to $B_\perp$, while spin couples to $|B|$. 
In fig.\ref{fig:figure4}(b), we compare the conductivity at $B_\perp=0$ for $B_\parallel=0$ and $B_\parallel=\SI{1.8}{T}$ (the maximum available in our system), where no significant effect is observed on $\sigma_{xx}$.
The lack of an in-plane magnetic field dependence is consistent with the presence of Ising SOC, which is expected to align spins out-of-plane. 
Therefore, for an in-plane magnetic field dependence in conductivity to occur, the Zeeman energy $\Delta E_\mathrm{Z}=2\mu_\mathrm{B} |B|$ would need to become comparable to the spin-orbit gap $\Delta_\mathrm{I}/2$, estimated to occur at $B > \SI{6.7}{T}$. 
In our experiments, the conductivity drops by nearly a factor of 2 at $B_\perp\approx \SI{50}{mT}$. 
This magnetic field corresponds to a Zeeman energy of only $\SI{6}{\micro eV}$, much smaller than disorder. Therefore, it is unlikely that the Zeeman effect could be responsible for the observed magnetoconductivity peaks.

\section{Discussion}

In this study we demonstrated that two type of SOC are present in spin-orbit proximitized BLG.
Despite the similar magnitudes of the two SOC terms, the band splitting at zero displacement field shows little dependence on the total density, indicating that the Ising SOC predominantly influences the splitting within the density range under investigation.
Our results align with previous observations of Ising superconductivity in WSe$_2$/BLG heterostructures \cite{holleis_ising_2023, zhang_enhanced_2023}, suggesting the potential for similar phenomena to occur in MoS$_2$/BLG systems.

Furthermore, we observed an insulating phase at $D=0$, leading to a non-monotonic electrical conductivity with respect to the displacement field.
Insulating phases with a similar displacement field dependence have been also observed in charge neutral suspended BLG \cite{weitz_broken-symmetry_2010}, albeit with an intrinsic SOC two orders of magnitude weaker than in our sample \cite{kane_quantum_2005, konschuh_theory_2012}.
While suspended BLG exhibits a gap at $B=0$ and $D=0$ \cite{velasco_transport_2012}, attributed to many-body correlations, our sample does not show this behavior [Fig.\ref{fig:figure4}(d)], suggesting a different underlying mechanism.
The absence of such correlated phases in hBN-encapsulated Bernal BLG suggests that dielectric and gate screening effects may reduce the relevance of correlation phenomena.
Thus, we conclude that SOC plays a crucial role in the emergence of the observed insulating phase at $D=0$.
This assertion aligns with findings by Island et al. \cite{island_spinorbit_2019}, who reported a comparable insulating phase in BLG fully encapsulated in WSe$_2$. 
While their explanation relied on SOC-driven band inversion, our observations suggest an alternative explanation, specifically single-particle SOC-induced gaps in spin-polarized bands in the absence of a global band gap [Fig.\ref{fig:figure1}(e)].
Our conclusion is supported by a detailed analysis of the SOC strength, a comparison between the band structure calculations and the displacement field dependence, as well as temperature dependent measurements.

While the zero magnetic field data are understood in terms of single-particle physics, we could not find a suitable theoretical model to describe the data at finite magnetic field.
We speculate that the non-monotonic magnetic field dependence of $\sigma_{xx}$ at $D = \pm D_\mathrm{c}$ originates from many-body effects at CN. 
Electron interactions, particularly strong near CN due to the lack of screening, were predicted to drive an instability towards an `excitonic insulator' phase, where carriers in valleys $K_{+}$ and $K_{-}$ exhibit strong particle-hole correlations \cite{zhang_spontaneous_2010, nandkishore_dynamical_2010, kharitonov_excitonic_2010, throckmorton_quantum_2014}.
Previous measurements, while showing promising results regarding gap opening at CN, were inconclusive.
This could be due to, among other reasons, a reduction in exchange interactions in the valley sector in the presence of spin degeneracy. 
In our system, with spin degrees of freedom polarized by SOC, carrier exchange responsible for the many-body physics at CN is expected to intensify.
If this interpretation holds true, the system described here could serve as a platform to explore various intriguing effects anticipated for excitonic phases, such as vortices, merons, and the Josephson effect for charge-neutral particles.

\textit{Note from the authors.}
While preparing our manuscript, we became aware of a related study by A. Seiler at al., who investigated the interplay between SOC and Coulomb interaction in WSe$_2$/BLG heterostructures, drawing conclusions on the phase diagram of SOC-proximitized BLG. 
It is remarkable that very similar data was obtained by two different groups, using a different TMD on bilayer graphene (MoS$_2$ by our research group and WSe$_2$ by Seiler et al.).

\newcommand{\beginMethods}{%
        \setcounter{table}{0}
        \renewcommand{\tablename}{Methods}
        \renewcommand{\thetable}{Tab.\arabic{table}}%
        \setcounter{figure}{0}
        \renewcommand{\figurename}{Methods}
        \renewcommand{\thefigure}{Fig.\arabic{figure}}%
        \setcounter{section}{0}
        \renewcommand\thesection{}
     }
     
\beginMethods

\section*{Methods}

\subsection{Sample fabrication}\label{met:Fab}

We initiate the fabrication of our devices by assembling the heterostructure using a polymer-based dry transfer technique. 
Each layer is obtained through mechanical exfoliation of bulk crystals onto silicon/silicon dioxide wafers. 
The heterostructure comprises, from top to bottom, hBN, bilayer graphene (BLG), three layers of MoS$_2$, hBN, and graphite. 
The finalized heterostructure is depicted in Extended Data~\ref{fig:SampleFab}(a).

The relative alignment of BLG with the MoS$_2$ layer is known to influence the strength of the SOC \cite{david_induced_2019, li_twist_2019}. 
While the maximum induced SOC is anticipated around $\SIrange{15}{20}{\degree}$, the SOC is most stable against small uncontrolled twist angle variations at $\SI{0}{\degree}$, ensuring better reproducibility. 
Therefore, during the fabrication process, we carefully align the edges of the MoS$_2$ and BLG flakes, resulting in potential relative alignments of $\SI{0}{\degree}$ or $\SI{30}{\degree}$. 
At $\SI{30}{\degree}$, the SOC proximity is expected to vanish, leading us to conclude that the relative angle in our sample is $\SI{0}{\degree}$.

Subsequently, the sample undergoes annealing in a hydro-argon atmosphere (H$_2$/Ar: 5\%/95\%) at $\SI{350}{\degree C}$ for 4 hours to remove polymer residues and enhance adhesion between the layers. The metallic top gate is defined using standard electron-beam lithography, followed by electron-beam evaporation (chromium/gold) and lift-off processes. The mesa is dry-etched using a reactive ion etching process with a CHF$_3$:O$_2$ mixture (40:4). An atomic force microscope image of the sample post-gate deposition and mesa etch is depicted in Extended Data~\ref{fig:SampleFab}(b).

In the final fabrication step, metallic edge contacts are deposited using electron-beam lithography, followed by electron-beam evaporation (chromium/gold) and lift-off processes. 
After resist development, we clean the contact area using an O$_2$ reactive ion etching process before metal deposition. This ensures the resulting contacts are ohmic and low resistive ($<\SI{1}{k\Omega}$). 
An optical image of the sample at the conclusion of the fabrication process is presented in Extended Data~\ref{fig:SampleFab}(c).

\subsection{Dual-gated device}

We employ a dual gate structure that allows for independent tuning of the charge carrier densities $n$ and displacement field $D$.
The density is defined as
\begin{equation}
    n = \frac{1}{e}\left( C_\mathrm{B} \vbg + C_\mathrm{T} \vtg \right) + n_0,
\end{equation}
and the displacement field is defined as
\begin{equation}\label{eq:D_field}
    D=\frac{1}{2} \left(C_\mathrm{B} \vbg - C_\mathrm{T} \vtg\right) + D_0,
\end{equation}
where $C_\mathrm{B}=\SI{36.7}{nF/cm^2}$ and $C_\mathrm{T}=\SI{78.2}{nF/cm^2}$ are the capacitance per area of the bottom and top gate, $\vbg$ and $\vtg$ are the voltages applied to the bottom and top gate.
Additionally, $n_0=-\SI{6.3E10}{cm^{-2}}$ and $D_0/\varepsilon_0=\SI{-46}{mV/nm}$ are offsets in the density and displacement field, respectively. These offsets are taken into account to compensate the asymmetries arising from factors such as the contact potential difference between hBN and MoS$_2$ \cite{gmitra_proximity_2017}.

\subsection{Measurements}

The measurements were performed in a pumped Helium-4 cryostat (for the temperatures above $\SI{1}{K}$) or in a dilution refrigerator with base temperature $<\SI{10}{mK}$ (estimated electronic temperature $\approx \SI{30}{mK}$).

The four-terminal resistance was measured with constant input current, by using a series resistor of $\SI{10}{M\Omega}$ or $\SI{100}{M\Omega}$, depending on the resistance of the sample. The input voltage was generated at a frequency of roughly \SI{31}{Hz} with a Lock-in amplifier.
The current amplitude ranged from \SIrange{1}{50}{nA}.

The bias spectroscopy measurements were done in a two terminal setup, where a DC voltage source was employed to generate the source-drain bias and a home-made voltage-to-current converter was used to detect the source-drain current.

\subsection{Tight-binding model}\label{met:Tight_binding}

To determine the band structure we employ a four-band effective tight-binding model for BLG in the basis $(A_1, B_1, A_2, B_2)$, where $A,B$ are the two atoms in the unit cell of a single graphene layer and their index represent the layer number \cite{mccann_electronic_2013}:
\begin{equation}\label{eq:Effective_H_BLG}
    H_{0} = \begin{pmatrix} 
        -U/2    & v_{0} \pi^\dagger  & -v_4\pi^\dagger     & v_3\pi\\ 
        v_0 \pi           & -U/2 +\Delta   & \gamma_1                & -v_4\pi^\dagger\\
        -v_4 \pi          & \gamma_1           & U/2+\Delta        & v_0\pi^\dagger \\
        v_3\pi^\dagger    & -v_4\pi            & v_0\pi& U/2
\end{pmatrix} ,
\end{equation}
where $\pi = \hbar(\xi k_x + ik_y)$, $\pi^\dagger = \hbar (\xi k_x -ik_y)$, $U$ is the inter-layer potential energy difference, $\Delta$ is an energy difference between dimer and non-dimer atoms, and $v_{j} = \frac{\sqrt{3}a}{2\hbar} \gamma_{j}$.
The parameters $\gamma_{j}$ are the Slonczewski–Weiss–McClure (SWM) parameters
given in Tab.~\ref{tab:SWM_params}.

\begin{table}[tb]
    \centering
    \caption{Values of the Slonczewski-Weiss-McClure (SWM) parameters in electron-Volt (eV). The experimental values are obtained from fits to infrared data (Ref.~\cite{zhang_determination_2008}). The second row provides the theoretical parameters obtained by \textit{ab initio} calculations based on local density approximation (LDA) \cite{jung_accurate_2014}. In this work we use the experimental values to calculate the band structure.}
\renewcommand{\arraystretch}{1.5}
\begin{tabular}{ cccccc } 
\toprule
SWM parameters & $\gamma_0$ & $\gamma_1$& $\gamma_3$ & $\gamma_4$ & $\Delta$\\
\hline
Exp.~\cite{zhang_determination_2008}    & 3.0  & 0.40  & 0.3   & 0.15  & 0.018 \\ 
Th.~\cite{jung_accurate_2014}           & 2.61 & 0.361 & 0.283 & 0.138 & 0.015 \\
\bottomrule
\end{tabular}
    \label{tab:SWM_params}
\end{table}

We include the extrinsic SOC given by equation~\eqref{eq:SO_hamiltonian}. The SOC lifts the spin degeneracy but does not mix states from different $K$-valleys. Therefore, the Hamiltonian becomes an $8\times 8$ matrix with the basis $(A_1 \uparrow, A_1 \downarrow, B_1\uparrow, B_1\downarrow,A_2 \uparrow, A_2 \downarrow, B_2\uparrow, B_2\downarrow)$. Since only layer 1 is in direct contact with the MoS$_2$ layer, the SOC is taken into account only in the top-left $4\times4$ block:
\begin{equation}
    H_\mathrm{SO} = \begin{pmatrix}
        H_\mathrm{SO}^\mathrm{L1} & 0\\ 0&0
    \end{pmatrix}
\end{equation}
The Ising and Rashba SOC components lead to the following $ H_\mathrm{SO}^\mathrm{L1}$ in matrix form:
\begin{equation}\label{eq:H_SO_Layer1}
    \begin{pmatrix} 
        \xi \frac{\Delta_\mathrm{I}}{2}     & 0   & 0    & -i\frac{\Delta_\mathrm{R}(\xi-1)}{2} \\ 
        0          & -\xi \frac{\Delta_\mathrm{I}}{2}    & i\frac{\Delta_\mathrm{R}(\xi+1)}{2}  & 0\\
        0         &-i\frac{\Delta_\mathrm{R}(\xi+1) }{2}           &\xi \frac{\Delta_\mathrm{I}}{2}   &0 \\
       i\frac{\Delta_\mathrm{R}(\xi-1)}{2}  & 0   & 0& -\xi \frac{\Delta_\mathrm{I}}{2}  
\end{pmatrix}.
\end{equation}

In the ordered basis $(A_1 \uparrow, A_1 \downarrow, B_1\uparrow, B_1\downarrow,A_2 \uparrow, A_2 \downarrow, B_2\uparrow, B_2\downarrow)$, the full Hamiltonian takes the form:

\begin{widetext}
\begin{equation}~\label{eq:appendix:fullHamiltonian}
   H= H_0 + H_\mathrm{SO} = \left(
\begin{array}{cccccccc}
 \frac{\Delta_I \xi }{2}-\frac{U}{2} & 0 & v_0 \pi^\dagger & -i\frac{\Delta_R (\xi - 1 )}{2}  & -v_4 \pi^\dagger & 0 & v_3 \pi & 0 \\
 0 & -\frac{\Delta_I \xi }{2}-\frac{U}{2} & i\frac{\Delta_R (\xi +1)}{2} & v_0 \pi^\dagger & 0 & -v_4 \pi^\dagger & 0 & v_3 \pi \\
 v_0 \pi & -i\frac{\Delta_R ( \xi +1)}{2} & \Delta +\frac{\Delta_I \xi }{2}-\frac{U}{2} & 0 & \gamma_1 & 0 & -v_4
   \pi^\dagger & 0 \\
 i\frac{\Delta_R (\xi -1)}{2} & v_0 \pi & 0 & \Delta -\frac{\Delta_I \xi }{2}-\frac{U}{2} & 0 & \gamma_1 & 0 &
   -v_4 \pi^\dagger \\
 -v_4 \pi & 0 & \gamma_1 & 0 & \Delta +\frac{U}{2} & 0 & v_0 \pi^\dagger & 0 \\
 0 & -v_4 \pi & 0 & \gamma_1 & 0 & \Delta +\frac{U}{2} & 0 & v_0 \pi^\dagger \\
 v_3 \pi^\dagger & 0 & -v_4 \pi & 0 & v_0 \pi & 0 & \frac{U}{2} & 0 \\
 0 & v_3 \pi^\dagger & 0 & -v_4 \pi & 0 & v_0 \pi & 0 & \frac{U}{2} \\
\end{array}
\right).
\end{equation} 
\end{widetext}

\subsubsection{Bands and Density of States}

The bands are then obtained by numerically diagonalizing $ H = H_0+H_\mathrm{SO}$. 
Each band is characterized by a band index $m$, which label the bands from the most negative ($m=0$) to the most positive ($m=7$) energies.

The density of states of band $m$ is given by
\begin{equation}\label{eq:DOS}
    \mathcal{D}_{m}(E) = \frac{1}{A} \sum_{\xi, \mb{k}} \delta(E-E_{m \xi, \mb{k}})
\end{equation}
where $\xi$ is the valley quantum number, and $\mb{k}$ is the wave vector. $A=L^2$ is the area in real space.
The delta function is approximated by a Gaussian function
\begin{equation}\label{eq:Gaussian}
   \delta(E-E_{m, \xi, \mb{k}}) \approx \frac{1}{\sqrt{2\pi} \epsilon} \exp\left(-\frac{(E-E_{m, \xi, \mb{k}})^2}{2\epsilon^2} \right), 
\end{equation}
with an energy broadening of $\epsilon<\SI{50}{\micro eV}$.
The band structure $E_{m,\xi,\mb{k} }$ is calculated on a grid in $k$ space with finite resolution $\Delta k\sim \SI{1E5}{m^{-1}}$.
Therefore the sum needs to be renormalized by the factor 
\begin{equation}\label{eq:DOS_norm}
    \left(\frac{\Delta k}{2\pi/L}\right)^2.
\end{equation}
Equations \eqref{eq:DOS}, \eqref{eq:Gaussian} and \eqref{eq:DOS_norm} yield the density of states of band $m$:
\begin{equation}
    \mathcal{D}_{m}(E) = \left(\frac{\Delta k}{2\pi}\right)^2 \sum_{\xi, \mb{k}} \frac{1}{\sqrt{2\pi} \epsilon} \exp\left(-\frac{(E-E_{m, \xi, \mb{k}})^2}{2\epsilon^2} \right).
\end{equation}
The total density of states is obtained by summing over the band index $m$.
 
The electron density is obtained by integrating over the conduction band ($m\geq 4$), while the hole density is obtained by integrating over the valence band ($m< 4$)
\begin{equation}
\begin{split}
    n_{e}(E_\mathrm{F})=\sum_{m=4}^{7}\int_{-\infty}^{E_\mathrm{F}} \mathcal{D}_{m}(E) \mathrm{d}E \\
    n_{h}(E_\mathrm{F})=\sum_{m=0}^{3}\int_{\infty}^{E_\mathrm{F}} \mathcal{D}_{m}(E) \mathrm{d}E.
\end{split}
\end{equation}
Out of the 8 bands, we only consider the four low energy bands ($2\leq m \leq 5$), $m=4,5$ for the conduction band and $m=2, 3$ for the valence band.
The total density is obtained by summation:
\begin{equation}
    n(E_\mathrm{F}) = n_e(E_\mathrm{F})-n_h(E_\mathrm{F}).
\end{equation}
According to our definition, hole doping corresponds to a negative densiy.

\subsection{Fitting routine}\label{met:Fitting}

To determine the spin-orbit parameters, we diagonalize the Hamiltonian for various SOC parameters and obtain the resulting band densities, as discussed above.
Then we calculate the standard deviation: 
\begin{equation}
    \sigma_\mathrm{std} = \sqrt{\frac{1}{N} \sum_{i=0}^{N}(\Delta n_{i} - \mu_{i})^2},
\end{equation}
where $N$ is the number of data points, $\Delta n_{i}$ is the experimental value of the density difference and $\mu_{i}$ is the expectation value. 
The resulting standard deviation is depicted in Extended Data~\ref{fig:App_StandardDeviation}. 
Notably, two minima with comparable standard deviations are observed for distinct parameter sets, marked by the red and blue dashed lines, respectively.
Despite the marginal disparity, we opt for the parameter combination corresponding to the minimum standard deviation (red dashed lines) to determine the SOC parameters, specifically $\Delta_\mathrm{I}=\SI{1.55\pm 0.1}{meV}$ and $\Delta_\mathrm{R}=\SI{2.5\pm 0.5}{meV}$.

\subsection{Interlayer screening}\label{met:Screening}

Due to the wave function polarization in BLG, interlayer screening effects reduce the effective interlayer potential.
Here, we consider the effect of interlayer screening and calcualte the conversion between the interlayer potential energy $U$ and the displacement field $D$ at charge neutrality.

The interlayer potential energy \cite{mccann_electronic_2013}
\begin{equation}
    U = \frac{e}{C_\mathrm{BLG}}\left(D-D_\mathrm{int}\right),
\end{equation}
is the result of an externally applied displacement field $D$ and an internal displacement field
\begin{equation}
    D_\mathrm{int} = \frac{e}{2}\Delta n_\mathrm{L}.
\end{equation}
Here, the screening density $\Delta n_\mathrm{L}=n_{1}-n_{2}$, $n_{i}$ being the electron density of layer $i$, represents the difference between the electron density of the two graphene layers.

The screening density is obtained by first determining the layer-resolved density of states
\begin{equation}
    \begin{split}
     \mathcal{D}_{i}(E)=& \left(\frac{\Delta k}{2\pi}\right)^2 \sum_{m, \xi, \mb{k}} \delta(E-E_{m \xi, \mb{k}}) \\ &\left[|\psi_{m,\xi,\mb{k}}^{A,i}|^2+|\psi_{m,\xi,\mb{k}}^{B,i}(\mb{k})|^2\right],   
    \end{split}
\end{equation}
where $\psi_{A,i}$ and $\psi_{B,i}$ are the wave functions of lattice site $A$ and $B$ on layer $i$.
Then the density is obtain by integration over the energy up to the Fermi energy $E_\mathrm{F}$:
\begin{equation}
    \Delta n_\mathrm{L}=\int_{-\infty}^{E_\mathrm{F}} \left[\mathcal{D}_1(E)-\mathcal{D}_2(E)\right]\mathrm{d}E .
\end{equation}
The lower limit of the integration is set to $E_\mathrm{min}=\SI{70}{meV}$, which is sufficiently large, such that $\Delta\mathcal{D}(E<E_\mathrm{min})\approx 0$ in the $U$ range considered here.

We calculate the screening density at charge neutrality point for interlayer potential energies $|U|\leq\SI{2}{meV}$, i.e. above $U_\mathrm{c}$, and calculate the corresponding displacement field according to 
\begin{equation}
    D = C_\mathrm{BLG}\frac{U}{e} + \frac{e\Delta n_\mathrm{L}}{2}.
\end{equation}
The result is shown in Extended Data~\ref{fig:app_Screened_U_vs_D} as black dots.
In this limited displacement field range, the function $U(D)$ is linear and its slope is reduced by \SI{42}{\percent} compared to the unscreened solution (dashed line).
The critical interlayer potential energy difference $U_\mathrm{c}=\Delta_{I}/2$ translates into a critical displacement field $D_\mathrm{c}=\SI{11.4}{mV/nm}$, matching the result of our experiment.

%


\section*{Acknowledgments}
We thank Thomas Weitz, Anna Seiler, and Patrick Lee for fruitful discussions.
We thank Peter Märki, Thomas Bähler, as well as the FIRST staff for their technical support.
We acknowledge support from the European Graphene Flagship Core3 Project, Swiss National Science Foundation via NCCR Quantum Science, and H2020 European Research Council (ERC) Synergy Grant under Grant Agreement 95154.
N.J. acknowledges funding from the International Center for Advanced Studies of Energy Conversion (ICASEC).
K.W. and T.T. acknowledge support from the JSPS KAKENHI (Grant Numbers 20H00354 and 23H02052) and World Premier International Research Center Initiative (WPI), MEXT, Japan.

\section*{Data availability}
Source data and analysis scripts associated with this study are made available via the ETH Research Collection (https://doi.org/10.3929/ethz-b-000662935).

\section*{Author contributions}
M.M., H.D., T.I., K.E. conceived and designed the experiments.
M.M., M.G., F.F, performed and analysed the measurements with inputs from H.D., J.G., and  M.N.. 
M.M. designed the figures with inputs from  C.T. and H.D..
M.M. and M.G. fabricated the device with inputs from J.G., M.N. and H.D..
A.P., N.J., L.L. provided theoretical support.
T.T., K.W. supplied the hexagonal boron nitride.
M.M wrote the manuscript with inputs from H.D..
All the coauthors mentioned above read and commented on the manuscript.

\clearpage
\newpage

\newcommand{\beginExtendedData}{%
        \setcounter{figure}{0}
        \renewcommand{\figurename}{Extended Data}
        \renewcommand{\thefigure}{Fig.\Roman{figure}}%
     }

\beginExtendedData

\makeatletter
\@fpsep\textheight
\makeatother

\begin{figure*}[h!]
    \centering
    \includegraphics{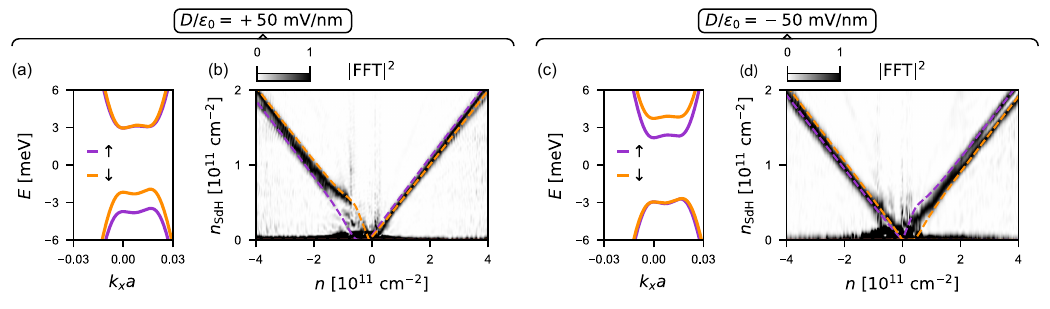}
    \caption{\textbf{Layer polarization at finite displacement field}. (a) Band structure, akin to Fig.~\ref{fig:figure1}(e), but at $D/\varepsilon_0=+\SI{50}{mV/nm}$. Positive displacement fields enhance the splitting in the valence band while suppressing it in the conduction band. This effect demonstrates wave function polarization in BLG, with valence band states polarized towards the layer adjacent to MoS$_2$, and conduction band states polarized away from MoS$2$.
    (b) FFT of the Landau fan measured at $D/\varepsilon_0=\SI{50}{mV/nm}$. The vertical axis is scaled to $n_\mathrm{SdH}=f~2e/h$. Dashed lines represent densities obtained from the band structure in (a).
    (c) and (d) correspond to (a) and (b), respectively, but for $D=-\SI{50}{mV/nm}$.}
    \label{fig:Layer_polarization}
\end{figure*}

\newpage

\begin{figure*}[h!]
    \centering
    \includegraphics{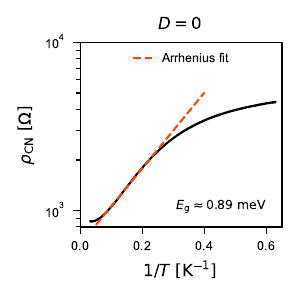}
    \caption{\textbf{Arrhenius plot.} Resistivity at charge neutrality as a function of the inverse temperature in a semi-logarithmic scale. The dashed line is a fit to the data with Arrhenius law in the temperature range $\SIrange{4}{10}{K}$.}
    \label{fig:Arrhenius}
\end{figure*}

\begin{figure*}[h!]
    \centering
    \includegraphics{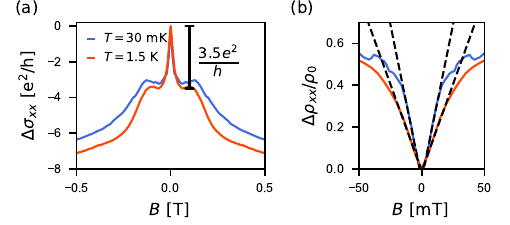}
    \caption{\textbf{Temperature dependence of the magnetoconductivity at CN and $D=D_\mathrm{c}$.} 
    (a) Magnetoconductivity $\Delta \sigma_{xx}=\sigma_{xx}(B)-\sigma_{xx}(0)$ as a function of magnetic field $B$ for two temperatures: $T=\SI{30}{mK}$ and $T=\SI{1.5}{K}$. The data show that the magnitude of the peak increases with temperature, in contrast to the suppression expected for WAL.
    (b) The same set of data but converted into magnetoresistance $\Delta\rho_{xx}/\rho_0$ plotted in a smaller magnetic field range.
    The dashed lines are linear fits to the data. These data show that the magnetoresistance is almost linear in this magnetic field range.
    }
    \label{fig:Tdependence_Peak}
\end{figure*}

\begin{figure*}[h!]
    \centering
    \includegraphics{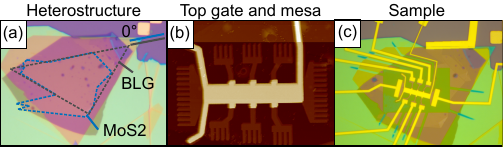}
    \caption{\textbf{Main steps of the fabrication process.} (a) Optical image of the heterostructure. The BLG and MoS$_2$ flakes are outlined in gray and blue, respectively.
    (b) Atomic force microscope image of the sample after gate deposition and mesa etch. 
    (c) Optical image of the sample at the end of the fabrication process.}
    \label{fig:SampleFab}
\end{figure*}

\begin{figure*}[h!]
    \centering
    \includegraphics{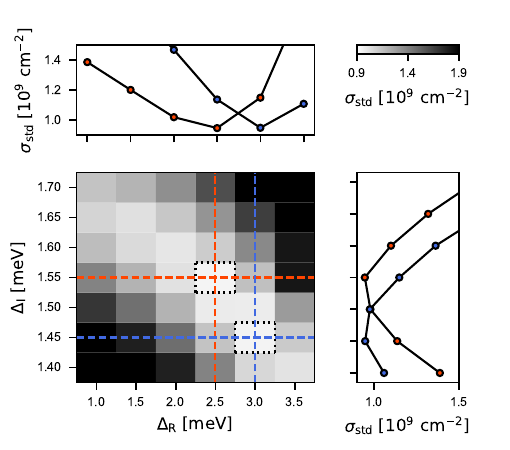}
    \caption{\textbf{Fitting routine}. Standard deviation of the fit to the curve $\Delta n - n$ presented in Fig.~\ref{fig:figure1}(f) in the main text. 
    The standard deviation consider both the electron and hole doping regimes.
    The top (side) panel shows horizontal (vertical) linecuts, highlighting that there are two almost equivalent minima of the standard deviation. The global minimum is shown by the red dashed line.}
    \label{fig:App_StandardDeviation}
\end{figure*}

\begin{figure*}[h!]
    \centering
    \includegraphics{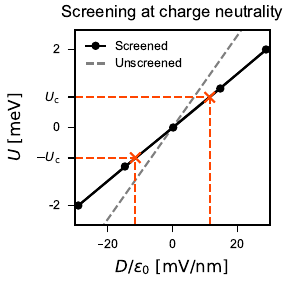}
    \caption{\textbf{Intelayer potential energy as a function of electric displacement field.} The gray dashed line shows the unscreened potential energy $U_0= eD/C_\mathrm{BLG}$, while the black line shows the result taking into account interlayer screening.
    The red dashed lines mark the critical displacement field $D_\mathrm{c}=\SI{11.4}{mV/nm}$ obtained for $U_\mathrm{c}=\SI{0.75}{meV}$.}
    \label{fig:app_Screened_U_vs_D}
\end{figure*}


\end{document}



\maketitle

\newcommand{\beginsupplement}{%
        \setcounter{table}{0}
        \renewcommand{\thetable}{SI\arabic{table}}%
        \setcounter{figure}{0}
        \renewcommand{\thefigure}{SI\arabic{figure}}%
     }
\beginsupplement

\section{Sample characterization}

\begin{figure}[h]
    \centering
    \includegraphics{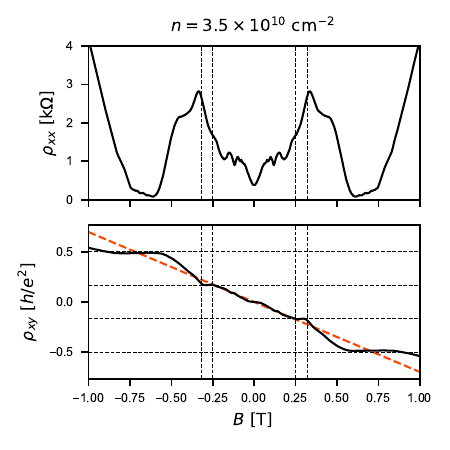}
    \caption{Longitudinal ($\rho_{xx}$) and Hall resistivity ($\rho_{xy}$) are plotted as a function of magnetic field ($B$) in the top and bottom panel, respectively. The measurement is performed in a dilution refrigerator at base temperature $T=\SI{30}{mK}$. }
    \label{fig:QH_plateus}
\end{figure}

The quality of our sample is evidenced by the magnetoresistance depicted in Fig.~\ref{fig:QH_plateus}. In the top panel, the longitudinal resistivity $\rho_{xx}$ exhibits Shubnikov–de Haas oscillations (SdHO) emerging below $B=\SI{100}{mT}$. The bottom panel displays the Hall resistivity $\rho_{xy}$, revealing well-defined quantum Hall plateaus at filling factor $\nu=2$, where $\rho_{xx}$ approaches zero. Interestingly, we also observe distinct plateaus at $\nu=6$, however, at this filling factor, $\rho_{xx}$ does not exhibit a minimum (indicated by the vertical dashed lines).

We estimate the transport mobility according to the Drude model 
\begin{equation}\label{eq:TransportMobility}
    \mu = \frac{1}{e n \rho_0},
\end{equation}
where $\rho_0=\SI{384}{\Omega}$ is the resistivity at $B=0$. The density $n=\SI{3.5E10}{cm^{-2}}$ is determined from the slope of $\rho_{xy}$ for $|B|<\SI{0.2}{T}$ (see red dashed line).
Substituting these values into equation~\ref{eq:TransportMobility} yields $\mu=\SI{4.65E5}{cm^2/(Vs)}$.
The mean-free-path is calculated using:
\begin{equation}
    \ell_e = \frac{h}{e^2} \frac{1}{\rho_0 \sqrt{2\pi n}},
\end{equation}
which gives $\ell_e\approx\SI{1.4}{\micro m}$. 
At a density of $n=\SI{4.95E10}{cm^{-2}}$, the mobility increases to $\SI{4.78E5}{cm^2/(Vs)}$, resulting in a mean-free-path of $\ell_e\approx\SI{1.7}{\micro m}$.
These values are comparable to the geometric dimensions of our device (width: $W=\SI{2}{\micro m}$, length: $L=\SI{14}{\micro m}$, and contact-to-contact distance: $L_\mathrm{C-C}=\SI{2.5}{\micro m}$), suggesting that the system operates in a quasi-ballistic regime.

\section{Comparison between models}

\begin{figure}[h!]
    \centering
    \includegraphics{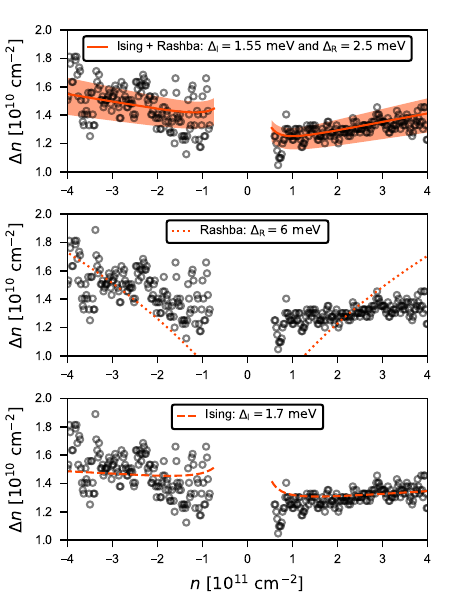}
    \caption{\textbf{Comparison between different fitting models.} (a) Fit presented in the main text, which includes both Ising and Rashba SOC terms (the parameters are shown in the legend). 
    (b) Fit that includes only the Rashba SOC. This model clearly deviates from the experimental data.
    (c) Fit that includes only Ising SOC. The fit reasonably capture the splitting observed in the experiment, but not the slope of $\Delta n(n)$.}
    \label{fig:App_Fit_Comparison}
\end{figure}

In Fig.~\ref{fig:App_Fit_Comparison}, we present three different fits: one considering both Ising and Rashba SOC (top panel), another considering only the Rashba SOC (central panel), and the last one considering only the Ising SOC (bottom panel).
These comparisons highlight that while a reasonable fit can be achieved when neglecting the Rashba SOC, the opposite is not true.
A fit including only the Rashba SOC yields poor agreement with the experimental data.
If we were to disregard the contribution of the Rashba SOC instead, we would need to compensate by increasing the Ising contribution.
However, such a model adjustment would fail to capture the observed slope in the experimental data, underscoring the importance of considering both contributions. 

The numerical outcome of our fitting process relies on the selection of the tight-binding coupling parameters for BLG.
These parameters influence the curvature of the bands, thereby affecting the conversion between density (the experimental parameter) and energy (the model parameter).
The curvature primarily depends on the intralayer coupling $\gamma_0$ and the nearest-neighbor interlayer coupling $\gamma_1$, while $\gamma_4$ determines the particle-hole asymmetry.
In our analysis, we utilize the experimentally determined coupling parameters listed in the first row of Tab.~1 in the Methods.
For comparison, we also conduct the fitting procedure using the theoretical values provided in the second row of the table.
This alternative approach yields $\Delta_\mathrm{I}=\SI{1.3\pm 0.1}{meV}$ and $\Delta_\mathrm{R}=\SI{2 \pm 0.5}{meV}$, which are in close agreement with the values reported in the main text.

\section{Layer-selective SOC}\label{app:Layer_selective_SOC}

Here, we discuss the layer polarization induced by the displacement field and its impact on the band structure of the proximitized BLG.
The displacement field can be converted into an on-site potential energy \cite{mccann_electronic_2013}:
\begin{equation}\label{eq:U_D_conversion}
    U = \frac{e}{C_\mathrm{BLG}}\left(D-e\frac{\Delta n_\mathrm{L}}{2}\right),
\end{equation}
where $C_\mathrm{BLG}\approx\SI{7.5}{\micro F/cm^2}$ \cite{sanchez_quantum_2012, rickhaus_electronic_2020} is the interlayer capacitance per area of BLG.
Here, $\Delta n_\mathrm{L}$ represents the electron density difference between the graphene layers, which screens the external electric field and reduces the effective potential energy $U$. 
Screening plays a crucial role, particularly near charge neutrality (CN), while its contribution diminishes at higher carrier densities ($n\gg\Delta n_\mathrm{L}$, where $\Delta n_\mathrm{L}$ depends on the applied displacement field). 
The role of screening is taken into account in Sec.~III, where we present data at CN, while it is neglected in the following discussion.

\begin{figure}[htb]
    \centering
    \includegraphics{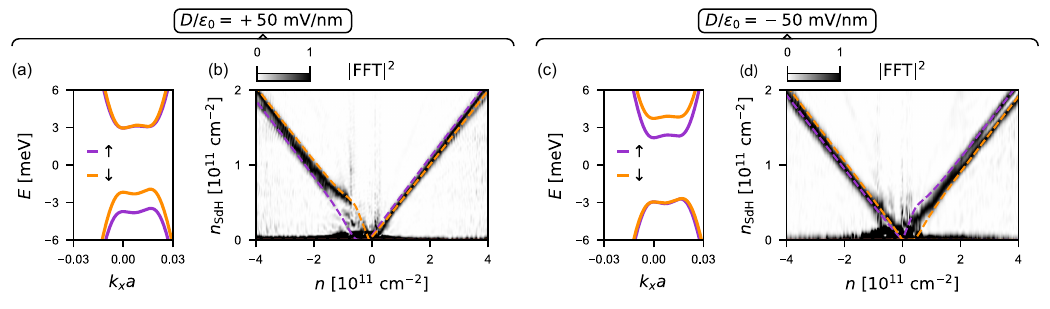}
    \caption{\textbf{Layer polarization at finite displacement field}. (a) Band structure, akin to Fig.~1(e) in the main text, but at $D/\varepsilon_0=+\SI{50}{mV/nm}$. Positive displacement fields enhance the splitting in the valence band while suppressing it in the conduction band. This effect demonstrates wave function polarization in BLG, with valence band states polarized towards the layer adjacent to MoS$_2$, and conduction band states polarized away from MoS$2$.
    (b) FFT of the Landau fan measured at $D/\varepsilon_0=\SI{50}{mV/nm}$. The vertical axis is scaled to $n\mathrm{SdH}=2ef/h$. Dashed lines represent densities obtained from the band structure in (a).
    (c) and (d) correspond to (a) and (b), respectively, but for $D=-\SI{50}{mV/nm}$.}
    \label{fig:figure2}
\end{figure}

The band structure of proximitized BLG is highly sensitive to changes in the on-site potential energy, as illustrated in Fig.~\ref{fig:figure2}(a) and (c).
A positive displacement field lowers the potential energy of electrons in the bottom layer of BLG, which is in direct contact with the MoS$_2$ layer.
Consequently, the electron wave function of the valence band states is drawn towards this layer, enhancing the effect of SOC proximity.
As a result, the spin-orbit splitting of the valence band increases.
In contrast, the wave function of the conduction band states is pushed away from the MoS$_2$ layer, leading to a suppression of the splitting.
The layer polarization of states is most pronounced at the band edges and gradually diminishes at higher energies.
Therefore, the splitting in the conduction band grows with increasing Fermi energy (i.e., electron density).
Reversing the polarity of the displacement field reverses the wave function polarization, causing the splitting to increase in the conduction band while being suppressed in the valence band.

The Fourier spectrum of the  SdHO measured under finite displacement fields is depicted in Fig.\ref{fig:figure2}(b) and (d) for $D/\varepsilon_0=+\SI{50}{mV/nm}$ and $D/\varepsilon_0=-\SI{50}{mV/nm}$, respectively.
Two notable distinctions between these spectra and those obtained at $D=0$ in Fig.~1(d) are observed: the splitting is either suppressed or enhanced for electron ($n>0$) or hole doping ($n<0$), depending on the polarity of the displacement field, aligning with expectations from band structure calculations.
This behavior corroborates the layer-selective SOC induced by substrate proximity \cite{khoo_on-demand_2017}.
Given the known SOC parameters from the $D=0$ data ($\Delta_\mathrm{I}=\SI{1.55}{meV}$, $\Delta_\mathrm{R}=\SI{2.5}{meV}$), we can directly compute the anticipated densities from the band structure, indicated by dashed lines in Fig.~\ref{fig:figure2}(b).
The model exhibits good agreement with the data, particularly at large carrier densities ($|n|>\SI{2E11}{cm^{-2}}$), while minor deviations are noticeable near CN.
These small deviations likely stem from interlayer screening, which is not accounted for in our model due to computational constraints.

\newpage

\section{Weak effect of the Rashba term on the low energy bands}\label{met:Rashba}

In the absence of spin-orbit coupling, the low energy bands near the $K\pm$ points primarily consist of the $A_1$ and $B_2$ orbitals \cite{mccann_electronic_2013}.
Despite the Rashba coupling $(\Delta_R = \SI{2.5}{meV})$ being stronger than the Ising SOC $(\Delta_I = \SI{1.55}{meV})$, the Rashba term has a weak effect on the low energy sector $(A_1 \uparrow, A_1 \downarrow, B_2\uparrow, B_2\downarrow)$.
This phenomenon arises because the Ising term is diagonal in the Hamiltonian, whereas the Rashba term is off-diagonal, coupling the low energy $A_1$ site with the $B_1$ site hosting bands in the high energy sector. 

\begin{figure*}[h!]
    \centering
    \includegraphics{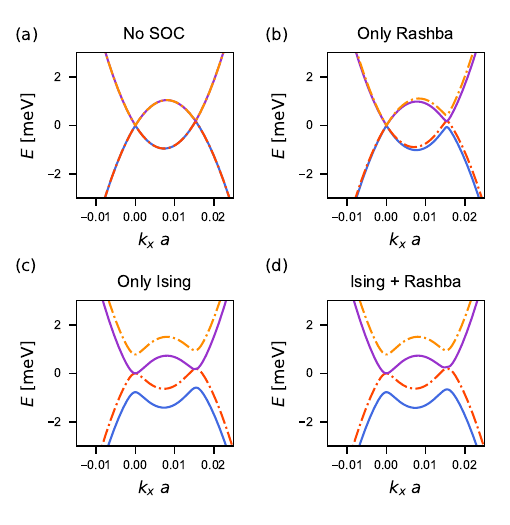}
    \caption{\textbf{Effects of SOC on the bandstructure near the $K_{+}$ point.} Here, $k_{x}$ denotes the momentum measured from the $K_{+}$ point. For small $k_{x}$, and for $U=0$, the red, blue, purple, and orange bands consist of superposition of the $A_1\downarrow, A_1\uparrow, B_2\downarrow$, and $B_2\uparrow$ orbitals. 
    (a) The low energy, spin-degenerate bandstructure obtained from $H_0$ (without any SOC) and $U=0$. 
    (b) The bandstructure of the full Hamiltonian $H$ in Methods with only Rashba SOC $\Delta_R = \SI{2.5}{meV}$, and no Ising SOC. 
    (c) The low energy bandstructure of $H$ with only Ising SOC $\Delta_I = \SI{1.55}{meV}$ and no Rashba SOC. 
    (d) The low energy bandstructure of $H$ with both Ising and Rashba SOC, $\Delta_I = \SI{1.55}{meV}$, $\Delta_R = \SI{2.5}{meV}$.}
    \label{fig:App_TightBinding}
\end{figure*}

The impact of the SOC terms is illustrated in Fig.~\ref{fig:App_TightBinding}. 
The low energy band structure without any SOC [Fig.~\ref{fig:App_TightBinding}(a)] and with only Rashba SOC [Fig.~ \ref{fig:App_TightBinding}(b)] are nearly indistinguishable.
On the other hand, the band structure with only Ising SOC [Fig.~\ref{fig:App_TightBinding}(c)] clearly shows a lifting of the band degeneracy. 
Upon including Rashba SOC [Fig.~ \ref{fig:App_TightBinding}(d)] alongside the Ising SOC, there is no significant modification of the bands, confirming the weak effect of the Rashba term.
Therefore, neglecting the Rashba term in theoretical calculations discussed in Sec.~III of the main text is justified.

\subsection{Layer polarization of the zeroth LL}\label{app:LP_zeroth_LL}

In the zeroth LL, there is a one-to-one correspondence between the valley and the layer degrees of freedom \cite{mccann_electronic_2013}. Consequently, applying a displacement field induces a valley polarization of the charge carriers. At finite magnetic field $B>0$ the Landau level degeneracy in each valley is $n_\xi=g eB/h$, where $g=4$ account for spin and orbital degrees of freedom. 
In case of full valley polarization, there will be an electrostatic energy stored in the BLG
\begin{equation}
   \epsilon = \frac{1}{2} \frac{Q^2}{C_\mathrm{BLG} }, 
\end{equation}
where $Q=e n_\xi A /2 $ ($A$ is the area of the capacitor) is the maximum amount of charges that can be layer polarized, as the zeroth Landau level is half filled at $\nu=0$. This energy is provided by the external displacement field $D$. 
The amount of charges that are polarized by the displacement field can be calculated using Gauss law
\begin{equation}
    \int \vec{D}\cdot\mathrm{d}\vec{A} = Q.
\end{equation}

In absence of SOC, the displacement field necessary to fully layer polarize the zeroth Landau level is
\begin{equation}
   D^*(B) = \frac{ge^2}{2h} B. 
\end{equation}
On the other hand, in the presence of Ising SOC, we need to take into account the displacement field $D_\mathrm{c}$, required to counteract the SO splitting, which yields
\begin{equation}\label{eq:PhaseBoundary}
   D^*(B) = \frac{ge^2}{2h} B + D_\mathrm{c}. 
\end{equation}

\begin{figure}[htb]
    \centering
    \includegraphics{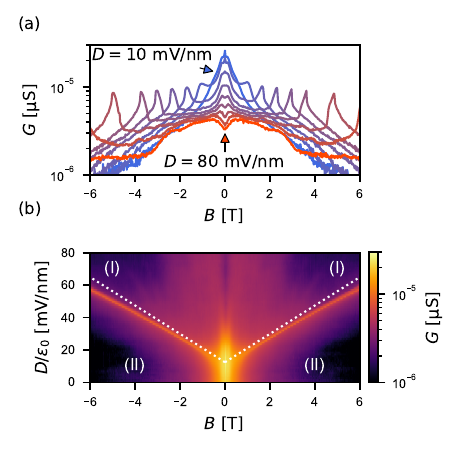}
    \caption{\textbf{Two terminal conductance at charge neutrality.} (a) Conductance as a function of of $B$ fo different displacement fields from $\SI{10}{mV/nm}$ to $\SI{80}{mV/nm}$. 
    (b) Conductance as a function of $B$ and $D$. The dotted lines are obtained by equation~\eqref{eq:PhaseBoundary}.}
    \label{fig:PhaseDiagram_TwoTermina}
\end{figure}

\clearpage 

\section{The gap in the canted antiferromagnetic phase}

\begin{figure}[htb]
    \centering
    \includegraphics{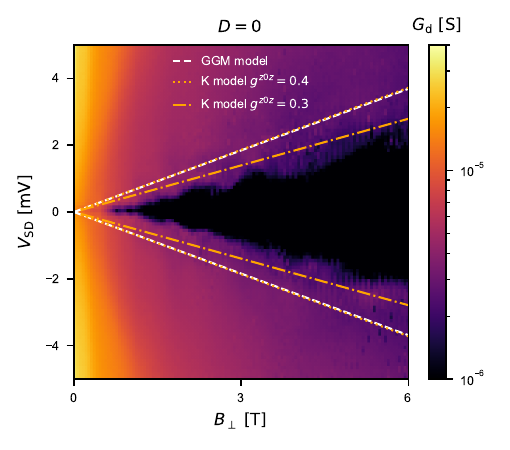}
    \caption{\textbf{Gap of the CAF phase.} Differential conductance $G_\mathrm{d}=\mathrm{d}I/\mathrm{d}V_\mathrm{SD}$ as a function of the source-drain bias voltage $V_\mathrm{SD}$ and magnetic field $B_\perp$.
    The white dashed line corresponds to the gap predicted for pristine BLG in Ref.~\cite{gorbar_energy_2010}, while the orange dotted line was obtained in Ref.~\cite{kharitonov_antiferromagnetic_2012} by adjusting the coupling parameter $g^{z0z}=0.4$ to the data of Velasco et al. \cite{velasco_transport_2012}.
    The coupling $g^{z0z}=0.3$ is adjusted to match the gap observed in our experiment (dashed-dotted line).}
    \label{fig:app_Gap_CAF}
\end{figure}

The presence of a gap at $D=0$ is demonstrated by the bias spectroscopy data presented in Fig.\ref{fig:app_Gap_CAF}. 
In this figure we present the differential conductance $G_\mathrm{d}=\mathrm{d}I/\mathrm{d}V_\mathrm{SD}$ as a function of the source-drain bias voltage $V_\mathrm{SD}$ and magnetic field $B_\perp$. 
The data reveal a range of bias voltage where the conductance is suppressed, indicating the presence of a gap ($E_\mathrm{g}$) in the spectrum. 
The gap shows a linear expansion with magnetic field, consistent with observations in pristine BLG \cite{velasco_transport_2012}. 

In our sample, $E_\mathrm{g}$ increases at a rate of \SI{0.39}{meVT^{-1}} [orange dashed-dotted line in Fig.\ref{fig:app_Gap_CAF}].
In the literature, two independent work done by Gorbar, Gusynin, and Miransky \cite{gorbar_energy_2010} and Kharitonov \cite{kharitonov_antiferromagnetic_2012} predicted a
slightly larger rate \SI{0.62}{meVT^{-1}} for pristine BLG (see white dashed and orange dotted lines). 
Nevertheless, we find a good agreement between the properties of the CAF phase in pristine BLG and Phase (II) observed in our sample.
 
Interestingly, in suspended bilayer graphene \cite{weitz_broken-symmetry_2010, velasco_transport_2012}, a gap of a few millielectron volts persisted even at $B=0$. 
However, our measurements do not resolve any gap at $B=0$. 
It is worth noting that conductance fluctuations lead to a soft gap, limiting our energy resolution. 
If a gap were present at $B=0$ due to interactions, it would likely be smaller than the energy scale dictated by disorder ($E_\mathrm{dis}\sim\SI{0.14}{meV}$).

\section{Weak-antilocalization}

We exclude that the magnetoconductance peaks observed in Fig. 3 of the main text arise from weak-antilocalization (WAL). 
Here, we fit the data with a WAL model to estimate the phase coherence length that would be attributed to such peaks and show that this phase coherence length would not meet the condition $\ell_\phi>\ell_e\sim\SI{1}{\micro m}$, which is required to observe the WAL effect.

The SOC in proximitized BLG is typically analysed with the model of McCann and Fal'ko \cite{mccann_symmetry_2012}, which predicts a quantum correction to the conductivity of graphene due to the presence of arbitrary SOC
\begin{equation}\label{eq:WAL}
    \delta \sigma(B) - \Delta\sigma(0) = -\frac{e^2}{2\pi h} \left[ F\left(\frac{B}{B_\phi}\right) - F\left(\frac{B}{B_\phi + 2B_{asy}}\right) -2F\left(\frac{B}{B_\phi +B_{asy} +B_{sym}}\right) \right],
\end{equation}
where
$$B_\mathrm{asy} = \frac{\hbar }{4De} \tau_{asy}^{-1} = \frac{\hbar}{4e\ell_{asy}^2}$$
$$B_\mathrm{sym} = \frac{\hbar }{4De} \tau_{sym}^{-1}  = \frac{\hbar}{4e\ell_{sym}^2}.$$

There are two type of SOC
\begin{itemize}
    \item the one that breaks $z\rightarrow -z$ symmetry (i.e. Bychkov-Rashba) leads to WAL
    \item the one that preserve $z\rightarrow -z$ symmetry (i.e. Kane-Mele type) leads to WL
\end{itemize}
As $B_\mathrm{sym}$ in the model only contributes to weak-localization (not observed in our experiment), we neglect this term in our analysis. 

\begin{figure}[htb]
    \centering
    \includegraphics{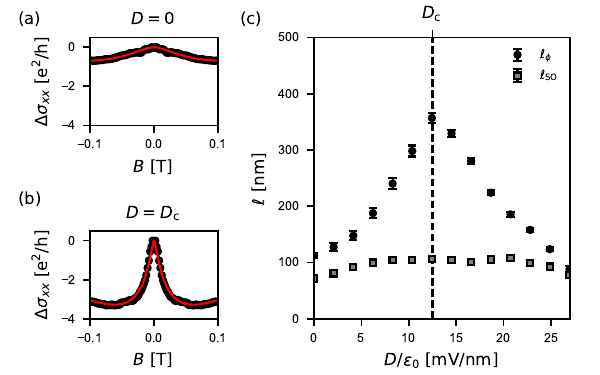}
    \caption{\textbf{Weak-antilocalization.}
    (a) and (b) shows the fit of the magnetoconductivity peak with Eq.\eqref{eq:WAL} for two exemplar displacement fields: $D=0$ and $D=D_\mathrm{c}$.
    (c) The fitting parameters $\ell_\phi$ and $\ell_\mathrm{SO}$ are plotted against the displaccement field.}
    \label{fig:app_WAL}
\end{figure}

This model predicts a maximum quantum correction to the conductivity of $0.5 e^2/h$.
Due to the larger magnitude of the peak observed in our experiment, we introduced an unrealistic scaling factor $\alpha$. The fitting parameters $\ell_\phi$ obtained from the fit depends on the choice of $\alpha$, as these parameters anti-correlates. 
However, a reasonable fit is obtained only for $\alpha\leq 16$.
Therefore, we fix $\alpha=16$ and fit the peak with Eq.~\eqref{eq:WAL} for different displacement fields, as shown in Fig.~\ref{fig:app_WAL}.
The fit yields a maximum coherence length of $\ell_\phi\approx\SI{360}{nm}$, therefore violating the condition $\ell_\phi>\ell_e\sim\SI{1}{\micro m}$.

\clearpage
\bibliographystyle{unsrt}